# A Framework for Pricing Schemes for Networks under Complete Information

Aditya Goyal[#1], Akanksha Tyagi[#2], Manisha Bhende[*3], Swapnil Kawade[#4]

[#]*Student & Department of Computer Engineering,*
*Padmashree Dr. D.Y. Patil Institute of Engineering & Technology,*
*Pimpri, Pune, Maharashtra, India*
[*]*Faculty & Department of Computer Engineering,*
*Padmashree Dr. D.Y. Patil Institute of Engineering & Technology,*
*Pimpri, Pune, Maharashtra, India*

*Abstract*— **The revenue maximization problem of service provider is considered and different pricing schemes to solve the above problem are implemented. The service provider can choose an apt pricing scheme subjected to limited resources, if he knows the utility function and identity of the user. The complete price differentiation can achieve a large revenue gain but has high implementation complexity. The partial price differentiation scheme to overcome the high implementational complexity of complete price differentiation scheme is also studied. A polynomial- time algorithm is designed for partial price differentiation scheme that can compute the optimal partial differentiation prices. The willingness of the users to pay is also considered while designing price differentiation schemes.**

## I. INTRODUCTION

Pricing is an important concern for management of networks. It also used for design and operation of networks. Many sophisticated pricing mechanisms to extract surpluses from the consumers and maximize revenue (or profits) for the providers have been proposed. A typical example is the optimal nonlinear pricing. In practice, however, it is often observed simple pricing schemes being deployed by the service providers. Typical examples include flat-fee pricing and (piecewise) linear usage-based pricing. The optimal pricing schemes derived in economics often have a high implementation complexity. Besides having a higher maintenance cost, complex pricing schemes are not "customer friendly" and discourage customers from using the services. The task of achieving the highest possible revenue often with complicated pricing schemes requires knowing the information (identity and preference) of each customer, which can be challenging in large scale communication networks.

The service provider wants to maximize its revenue by setting the right pricing scheme to induce desirable demands from users. Since the service provider has a limited total resource, it must guarantee that the total demand from users is no larger than what it can supply. The details of pricing schemes depend on the information structure of the service provider. Under complete information, since the service provider can distinguish different groups of users, it announces the pricing and the admission control decisions to different groups of users.

In this paper, the optimal usage-based pricing problem in a resource-constrained network with one profit-maximizing service provider and multiple groups of surplus-maximizing users is studied. In wireless communication networks, however, the usage-based pricing scheme seems to become increasingly popular due to the rapid growth of wireless data traffic.

## II. RELATED WORK

The maximum revenue that can be achieved by a monopolistic service provider under complete network information has been studied in [1]. The authors proposed two pricing schemes with incomplete information, and showed that by properly combining the two schemes there would be very small revenue loss in a two-group case while maintaining the incentive compatibility.

A model to study the important role of time-preference in network pricing has been presented in [2]. In the model presented, each user chooses his access time based on his preference, the congestion level, and the price that he would be charged. Without pricing, the "price of anarchy" (POA) can be arbitrarily bad. The authors then derived a simple pricing scheme to maximize the social welfare. From the SP's viewpoint, the authors considered the revenue-maximizing pricing strategy and its effect on the social welfare. The authors showed that if the SP can differentiate its prices over different users and times, the maximal revenue can be achieved, as well as the maximal social welfare. However, if the SP had insufficient information about the users and can only differentiate its prices over the access times, then the resulting social welfare, especially when there are many low-utility users, can be much less than the optimum. Otherwise, the difference is bounded and less significant.

The results on price-based discrimination for bandwidth allocation in wire-line communication networks were presented in [3]. In general, the problem of mechanism design for resource allocation was very complex, and the focus was on studying simple mechanisms that show promise of widespread adoption in the arena of Internet pricing. The objective was to study the revenue efficiency of single and multi-class pricing schemes as compared to the maximum possible revenue.

In particular, the focus was on flat entry fees as the simplest pricing rule. A lower bound for the ratio between the revenue from this pricing rule and maximum revenue, which the author referred to as the Price of Simplicity was presented. The characterizations of types of environments that lead to a low Price of Simplicity was done and it was shown that the loss of revenue from using simple entry fees was small in a range of environments.

In the communication network pricing literature, it is the linear pricing schemes that have been largely adopted as the means of controlling network usage or generating profits for network service providers. In [4], the authors extended the framework mentioned above to non-linear pricing and investigated optimal nonlinear pricing policy design for a monopolistic service provider. The problem was formulated as an incentive-design problem, and incentive (pricing) policies were obtained for a many-users regime, which enabled the service provider to approach arbitrarily close to Pareto-optimal solutions.

### III. EXISTING SYSTEM

A system model of charging, routing and flow control, where the system comprises both users with utility functions and a network with capacity constraints has been described by Frank Kelly[5]. Standard results from the theory of convex optimization show that the optimization of the system may be decomposed into subsidiary optimization problems, one for each user and one for the network, by using price per unit flow as a Lagrange multiplier that mediates between the subsidiary problems.

TCP variants have recently been reverse-engineered to show that they are implicitly solving this problem, where the source rate vector $x \geq 0$ is the only set of optimization variables, and the routing matrix R and link capacity vector c are both constants in [9].

$$\text{maximize} \sum U_s(x_s)$$
$$\text{subject to } R_x \leq c$$

Utility functions $U_s$ are often assumed to be smooth, increasing, concave, and dependent on local rate only, although recent investigations have removed some of these assumptions for applications where they are invalid.

Flat-fee pricing: For Internet service providers, flat rate is access to the Internet at all hours and days of the year (linear rate) and for all customers of the telco operator (universal) at a fixed and cheap tariff. Flat rate is common in broadband access to the Internet in the USA and many other countries.

A charge tariff is a class of linear rate where the user is charged on the basis of uploads and downloads (data transfers) and hence differs from the flat rate system. Some GPRS / data UMTS access to the Internet in some countries of Europe has no flat rate pricing, following the traditional "metered mentality". Because of this, users prefer using fixed lines (with narrow or broadband access) to connect to the Internet.

### IV. PROPOSED METHODOLOGY

The partial price differentiation problem includes complete price differentiation scheme and single pricing scheme as special cases. The optimal solution to partial price differentiation problem is found out. The differentiation gain and the effective market size are the two important factors behind the revenue increase of price differentiation schemes.

A network with a total amount of S limited resource is considered. The resource can be in the form of rate, bandwidth, power, time slot, etc. The monopolistic service provider allocates the resource to a set $I = \{1, \ldots, I\}$ of user groups [8]. Each group $i \in I$ has $N_i$ homogeneous users with the same utility function:

$$u_i(s_i) = \Theta_i \ln(1 + s_i)$$

Where $s_i$ is the allocated resource to one user in group i and $\Theta_i$ represents the willingness to pay of group i.

It is assumed that $\Theta_1 > \Theta_2 > \ldots > \Theta_I$. Since the service provider has a limited total resource, it must guarantee that the total demand from users is no larger than what it can supply. The details of pricing schemes depend on the information structure of the service provider.

The correspondence between the service provider and user can be described as follows:

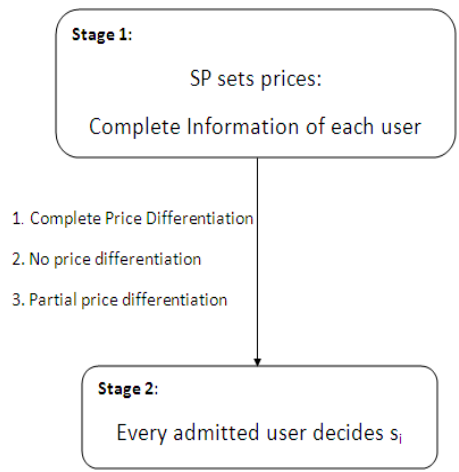

Fig. 1. General Proposed Model

The service provider declares the pricing schemes in Stage 1 and users interact with their demands in Stage 2. The users demand to maximize their surplus by optimizing their claim according to the pricing scheme. The service provider maximizes his revenue by making available a right pricing scheme to users.

It is considered that the service provider has the complete information of the user. The service provider can choose from complete price differentiation scheme, the single pricing scheme, and the partial price differentiation scheme.

*A. Complete Price Differentiation*

The service provider knows the utility and the identity of each user; it is possible to maximize the revenue by charging a different price to each group of users.

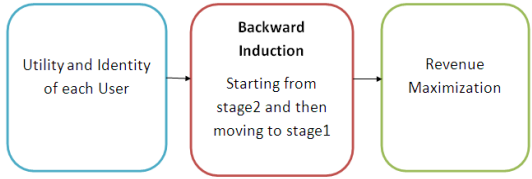

Fig. 2 Complete Price Differentiation

Algorithm:
Step 1: Solve users' maximization problem which leads to unique optimal demand.
Step 2: In stage 1, the service provider maximizes his revenue by choosing a price $P_i$ and number of users 'n' for each group I, subject to total resource 'S'.
Step 3: Perform Complete Price Differentiation scheme by charging each group by different price.

*B. Single Pricing*

In this scheme, the service provider charges a single price to all groups of users. This scheme may suffer a considerable revenue loss.

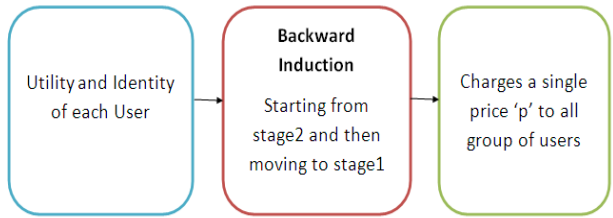

Fig. 3 Single Pricing

Algorithm:
Step 1: Single pricing follows the same approach as complete pricing scheme. A solution is obtained which shares a similar structure as complete price differentiation.
Step 2: There exists an optimal solution of this scheme that satisfies following conditions:
- All users are admitted: $n^*_I = N_i$ for all $i \in I$[8].
- There exists a price $p^*$ and a group index threshold $K^{sp} \leq I$ such that only the top $K^{sp}$ groups of users receive positive resource allocations [8].

*C. Partial Price Differentiation*

The service provider offers only a few pricing plans for the entire users population; it is termed as the *partial price differentiation* scheme. The clusters are defined. Each cluster is a set of groups which have been charged same price. The partial price differentiation problem is solved in three levels.

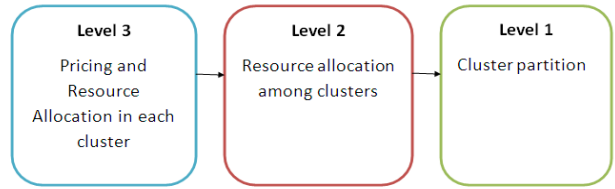

Fig. 4 Partial Price Differentiation

Algorithm:
Step 1: Pricing and resource allocation in each cluster: For a fix partition α and a cluster resource allocation, focus on pricing and resource allocation problem within each cluster.
Step 2: Resource allocation among clusters: For a fix partition α allocate resources among clusters.
Step 3: Cluster Partition: Cluster partition problem is solved.

V. MATHEMATICAL MODEL

*A. Complete Price Differentiation*

This method can be carried out in two stages as given below:

1) *User's Surplus Maximization Problem:*

If a user in group i has been admitted into the network and offered a linear price $p_i$, then it solves the following surplus maximization problem [8],

Maximize $\quad u_i(s_i) - p_i s_i$
$\quad\quad\quad\quad s_i \geq 0$

2) *Service Provider's Pricing and Admission Control Problem:*

CP: maximize $\quad \sum n_i p_i s_i$
$\quad p \geq 0, s \geq 0, n \quad i \in G$
subject to $s_i = (\Theta_i / p_i - 1)^+, i \in G$,
$n_i \in \{0, \ldots, N_i\}, i \in G$,
$\sum n_i s_i \leq S$
$i \in G$

*B. Single Pricing*

This problem can be formulated as given below [8]:
SP: $\quad$ maximize $\quad p \sum n_i s_i$
$\quad\quad p \geq 0, n \quad\quad i \in G$
subject to $s_i = (\Theta_i / p_i - 1)^+, i \in G$,
$n_i \in \{0, \ldots, N_i\}, i \in G$,
$\sum n_i s_i \leq S$
$i \in G$

Here the service provider charges a single price p to all groups of users.

*C. Partial Price Differentiation*

The Partial Price differentiation (PP) problem is formulated as follows [8].

PP: maximize $\sum_{i \in G} n_i p_i s_i$
$n_i, p_i, s_i, p^j, a_i^j$

subject to $s_i = (\Theta_i / p_i - 1)^+$, $\forall i \in G$,
$n_i \in \{0, \dots, N_i\}$, $\forall i \in G$,
$\sum_{i \in G} n_i s_i \leq S$

The service provider can choose the price charged to each group.

## VI. ANALYSIS MODEL

The system consists of two modules as shown in Fig. 5.

- **User Module**

The User has to register first and only then he can use the product.

- **Service Provider Module**

Service Provider selects a pricing scheme to maximize his revenue.

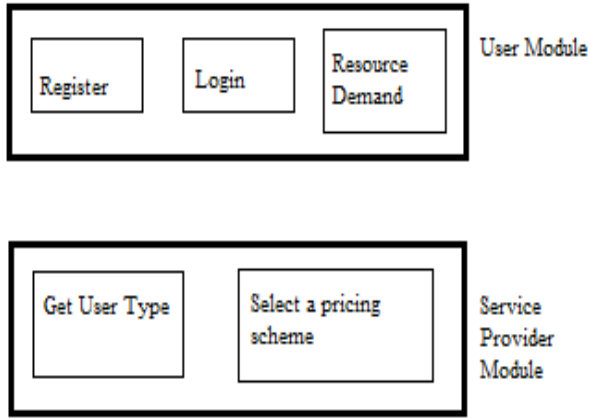

Fig. 5 System Architecture

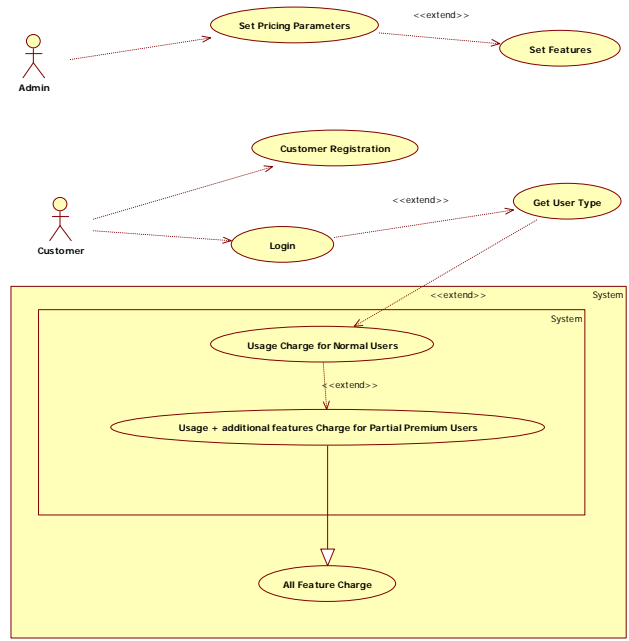

Fig. 6 System Analysis Model

## VII. APPLICATIONS

- The Omaha based Network Pricing, Mitec, has a variety of pricing options that target personal, business and corporate users. They offer Web development solutions and a variety of different access speeds. For users seeking "personal solutions," for example, Mitec offers them an option of a flat-fee account with unlimited access for $19.95 (flat-fees also help in gaining market share), or a tiered account with a flat-fee of $9.95 for the first 20 hours and a $1 additional charge for each hour thereafter. For the family on the Internet, Mitec offers unlimited access for $24.95 with five separate email accounts.

- AT&T shifted to usage-based wireless data plans in June 2010 [6]. AT&T announced two new wireless data plans based on the amount of data subscribers use. The change spelled the end of unlimited wireless data use for new customers and likely higher charges for existing customers who use more than 2GB of data per month for activities such as watching videos and online gaming.

- Verizon followed up with similar plans (like AT&T) in July 2011. Similar usage-based pricing plans have been adopted by major Chinese wireless service providers including China Mobile and China UniCom.

VIII. CONCLUSION

The revenue-maximizing problem for a monopoly service provider under complete network information is studied. The focus is to investigate the trade-off between the total revenue and the implementational complexity (measured in the number of pricing choices available for users) under complete information. The partial price differentiation is the most general one among the three pricing differentiation schemes that have been proposed (*i.e.,* complete, single, and partial), and includes the other two as special cases. An algorithm that computes the optimal partial pricing scheme in polynomial time, and numerically quantizes the trade-off between implementational complexity and total revenue has been designed.